\documentclass[eqsecnum,twocolumn,showpacs,preprintnumbers,amssymb,aps]{revtex4}
\usepackage{graphicx}
\usepackage{dcolumn}
\usepackage{bm}
\usepackage{latexsym,epsfig}

\begin{document}

\preprint{\today}

\title{Electron correlation effects in the dipole polarizabilities of the ground states of Be, Mg, Ca, Sr, Ba and Yb} 
\vspace{0.5cm}

\author{B. K. Sahoo \protect \footnote[1]{E-mail: bijaya@mpipks-dresden.mpg.de}}
\affiliation{Max-Planck Institute for the Physics of Complex Systems \\ N\"othnitzer Stra{\ss}e 38, D-01187 Dresden, Germany}
\author{B. P. Das \protect \footnote[2]{E-mail: das@iiap.res.in}}
\affiliation{Non-Accelerator Particle Physics Group, Indian Institute of Astrophysics, Bangalore-560034, India}
\date{\today}
\vskip1.0cm

\begin{abstract}
\noindent
We investigate the role of electron correlation effects in the electric dipole
polarizabilities of the ground states of the alkaline earth and
ytterbium atoms by employing the relativistic coupled-cluster (RCC) theory.
These effects are incorporated via the residual Coulomb interaction to
all orders in the RCC singles and doubles approximation. The perturbed wavefunctions used in the calculations of the polarizabilities are obtained by directly solving the first order perturbed RCC equations, thereby avoiding the sum-over-states approach. Our results are compared with other calculations and available experimental data.
\end{abstract} 

\pacs{31.15.Ar,31.15.Dv,31.25.Jf,32.10.Dk}
\keywords{Ab initio method, polarizability}
\maketitle

\section{Introduction}
 A knowledge of electric dipole polarizabilities is necessary in many areas of
physics and chemistry. In particular it is required in studies
of collisions involving atoms and molecules \cite{miller1}. In recent years,
the pre-eminent role of polarizabilities in the determination of inter atomic
interactions has assumed special significance in the context of research on
ultra-cold atoms \cite{pethick}.

Calculations of atomic polarizabilities have come a long way since the
classic work of Dalgarno and Lewis \cite{dalgarno}. Following
a series of calculations using the coupled Hartree-Fock method (see for
example, \cite{dalgarno1}), a number of state-of-the art methods
including the coupled-cluster (CC) theory have been used to calculate
atomic and molecular polarizabilities \cite{kundu,stanton,kallay}.
A few calculations of the polarizabilities of heavy atomic systems have
been performed in the past few years using the linearized \cite{safronova}
as well as the non-linearized \cite{bijaya1} relativistic CC theory.
These are based on approaches that sum over a set
of intermediate states. In contrast, we have obtained the first order
perturbed wavefunction by solving the first order perturbed CC
equation and used it to obtain the dipole polarizabilities of the
closed-shell alkaline earth atoms and ytterbium (Yb). Information on these
quantities is useful for the frequency standards experiments that have
been proposed for Mg \cite{porsev}, Ca \cite{degenhardt}, Sr \cite{boyd} and 
Yb \cite{barber} as well as the search
for the electric dipole moment in Yb \cite{takahashi}.

There has been considerable interest in accurate calculations of the
dipole polarizabilities of alkaline earth atoms and Yb. Sadlej et al
have calculated these quantities for  Ca, Sr and Ba using a
quasi relativistic approach but their treatment of correlation is
rigorous \cite{andrzej}. Their calculations have been carried
out at the finite order many-body perturbation theory and CC
levels. Porsev and Derevianko have performed calculations of the dipole
polarizabilities of Mg, Ca, Sr and Yb by a hybrid approach involving 
many-body perturbation theory and the configuration interaction method 
\cite{porsev}. A calculation of the dipole polarizability
of Yb based on the time dependent density functional theory (TDDFT) has been
reported recently \cite{chu}.

\section{Theory and Method of Calculations}
In a DC electric field ${\bf E}= \cal E {\bf \hat{z}}$, the energy shift 
$\Delta E$ of the ground state $|\Psi^{(0)}(\gamma, J_0,M_0)>$ with the parity 
eigenvalue
$\gamma$ and angular momentum $J_0$ and its azimuthal value $M_0$ is given by
\begin{eqnarray}
\Delta E = -\frac{1}{2} \alpha {\cal E}^2,
\end{eqnarray}
where $\alpha$ is the static polarizability and can be defined 
as
\begin{eqnarray}
\alpha = -2 \sum_{I} \frac {|\langle \Psi^{(0)}(\gamma,J_0,M_0)|D_z| \Psi^{(0)}(\gamma',J_I,M_I) \rangle|^2} {E_0 - E_I} ,
\label{eqn23}
\end{eqnarray}
where $D_z$ is the $z$th component of the electric dipole operator 
depending upon the applied field ${\bf E}$, subscript $I$ denotes the
intermediate states and $J_I$, $M_I$ are their angular momentum quantum
numbers, $\gamma$ and $\gamma'$ are parity quantum numbers for states of 
opposite parity and $E_0$ and $E_I$ are the energy values of
the ground state and the corresponding intermediate states $I$, respectively.

In a more explicit form, the above expression can be written as
\begin{widetext}
\begin{eqnarray}
\alpha &=& -2 \sum_{I} \frac {\langle \Psi^{(0)}(\gamma,J_0,M_0)|D_z| \Psi^{(0)}(\gamma',J_I,M_I) \rangle \langle \Psi^{(0)}(\gamma',J_I,M_I)|D_z| \Psi^{(0)}(\gamma,J_0,M_0)\rangle} {E_0 - E_I} \nonumber \\
 &=& - \sum_{I} \frac {\langle \Psi^{(0)}(\gamma,J_0,M_0)|D_z| \Psi^{(0)}(\gamma',J_I,M_I) \rangle \langle \Psi^{(0)}(\gamma',J_I,M_I)|D_z| \Psi^{(0)}(\gamma,J_0,M_0)\rangle} {E_0 - E_I} \nonumber \\ && - \sum_{I} \frac {\langle \Psi^{(0)}(\gamma',J_I,M_I)|D_z| \Psi^{(0)}(\gamma,J_0,M_0) \rangle \langle \Psi(\gamma',J_I,M_I)|D_z| \Psi(\gamma,J_0,M_0)\rangle} {E_0 - E_I} \nonumber \\
 &=& - \langle \Psi^{(0)}(\gamma,J_0,M_0)|D_z | \Psi^{(1)}(\gamma',J_0,M_0)\rangle + \langle \Psi^{(1)}(\gamma',J_0,M_0) | D_z | \Psi^{(0)}(\gamma,J_0,M_0)\rangle \nonumber \\
 &=& - 2 \langle \Psi^{(0)}(\gamma,J_0,M_0)|D_z | \Psi^{(1)}(\gamma',J_0,M_0)\rangle \nonumber \\
 &=& - \langle \Psi(\gamma,J_0,M_0)|D_z | \Psi(\gamma,J_0,M_0)\rangle \nonumber \\
\label{eqn6}
\end{eqnarray}
\end{widetext}
where we define $| \Psi^{(1)}(\gamma',J_0,M_0)\rangle$ as the first order
correction to the original unperturbed wavefunction $ |\Psi^{(0)}(\gamma,J_0,M_0)\rangle$ 
due to the operator $D_z$ and hence the total wavefunction in the presence of
an external DC electric field is given by
\begin{eqnarray}
|\Psi(\gamma,J_0,M_0)\rangle &=& |\Psi^{(0)}(\gamma,J_0,M_0)\rangle + |\Psi^{(1)}(\gamma',J_0,M_0)\rangle . \nonumber \\
\label{eqn7}
\end{eqnarray}

Our aim in this work is to obtain the exact wavefunction $|\Psi(\gamma,J_0,M_0)\rangle$ by calculating both $|\Psi^{(0)}(\gamma,J_0,M_0)\rangle$ and $|\Psi^{(1)}(\gamma,J_0,M_0)\rangle$ wavefunctions using an approach which
can rigorously incorporate the relativistic and correlation effects . In 
other words, we would like to obtain $|\Psi^{(1)}(\gamma',J_0,M_0)\rangle$ for 
the Dirac-Coulomb (DC) wavefunction $|\Psi^{(0)}(\gamma,J_0,M_0)\rangle$ as
the solution of the following equation
\begin{widetext}
\begin{eqnarray}
(\text{H}_0^{(\text{DC})} - E_n^{(0)})|\Psi^{(1)}(\gamma',J_0,M_0)\rangle = (E_0 - \text{H}_{\text{int}}) |\Psi^{(0)}(\gamma,J_0,M_0)\rangle ,
\label{eqn8}
\end{eqnarray}
\end{widetext}
where $\text{H}_0^{(\text{DC})}$ and $\text{H}_{\text{int}}$ correspond 
to the DC Hamiltonian and interaction due to $D_z$, respectively.

Using the CC ansatz, we express wavefunctions $|\Psi(\gamma,J_0,M_0)\rangle$ by \cite{cizek}
\begin{eqnarray}
|\Psi(\gamma,J_0,M_0)\rangle &=& e^T |\Phi_0(\gamma,J_0,M_0)\rangle ,
\label{eqn9}
\end{eqnarray}
where $|\Phi_0(\gamma,J_0,M_0)\rangle$ are the Dirac-Fock (DF) wavefunctions 
determined using the mean-field approximation and $T$ are the electron 
excitation  operators from the corresponding DF states.

 To obtain both the unperturbed and perturbed wavefunctions of Eq. (\ref{eqn7})
separately, we divide the excitation operators $T$ as
\begin{eqnarray}
T &=& T^{(0)} + \lambda T^{(1)} 
\label{eqn11}
\end{eqnarray}
where $T^{(0)}$ and $T^{(1)}$ are the all order excitation operator 
of the relativistic coupled-cluster (RCC) method and its first order    
correction arising in the presence of $D_z$, respectively.  $\lambda$ represents
the perturbation parameter . More explicitly, we can 
write the unperturbed and perturbed wavefunctions in terms of the RCC excitation
 operators as
\begin{eqnarray}
|\Psi^{(0)}(\gamma,J_0,M_0) \rangle &=& e^{T^{(0)}} |\Phi_0(\gamma,J_0,M_0)\rangle , \\
|\Psi^{(1)}(\gamma,J_0,M_0) \rangle &=& e^{T^{(0)}} T^{(1)} |\Phi_0(\gamma,J_0,M_0)\rangle ,
\label{eqn12}
\end{eqnarray}
where the exponential function of $T^{(1)}$ reduces to the linear term since we 
have considered only one order of $D_z$ operator in Eq. (2.5).

In the present work, we have considered all possible single and double
excitations (CCSD approach) in the calculations. We obtain first
the unperturbed $T^{(0)}$ amplitudes by solving the usual (R)CC equations, 
then these amplitudes are used to determine the $T^{(1)}$ amplitudes.
The corresponding equations to solve these amplitudes are given by
\begin{eqnarray}
\langle \Phi_0^{*} | \overline{\text{H}_N^{(\text{DC})}} | \Phi_0 \rangle &=& 0 \\
\langle \Phi_0^{*} | \overline{\text{H}_N^{(\text{DC})}} T^{(1)} | \Phi_0 \rangle &=&  -\langle \Phi_0^{*} | \overline{\text{H}_{\text{int}}} | \Phi_0 \rangle 
\end{eqnarray}
where the subscript $N$ represents normal order form of the operators and
we have defined $\overline{\text{H}}= e^{-T^{(0)}} \text{H} e^{T^{(0)}} = (\text{H} e^{T^{(0)}})_{con}$, with the subscript '$con$' representing connected terms.

The polarizabilities are determined by evaluating the following expression
\begin{eqnarray}
\alpha &=& \frac {\langle \Phi_0 | e^{T^{\dagger}} D e^T | \Phi_n \rangle } { \langle \Phi_0 | e^{T^{\dagger}}  e^T | \Phi_0 \rangle } \nonumber \\
&=& \frac { \langle\Phi_0| (T^{(1)^{\dagger}}\overline{\text{D}^{(0)}}+\overline{\text{D}^{(0)}}T^{(1)}) |\Phi_0>}{ \langle\Phi_0|e^{T^{(0)^{\dagger}}} e^{T^{(0)}}| \Phi_0 \rangle } , \ \ \ \
\label{eqn19}
\end{eqnarray}
where for computational simplicity we define $\overline{\text{D}^{(0)}}=e^{T^{(0)^{\dagger}}} \text{D} e^{T^{(0)}}$. We compute these terms after expressing
them as effective one-body and two-body terms using the generalized Wick's theorem
\cite{lindgren}. 

\section{Results and Discussions}
In Table \ref{tab1}, we present our results of electric dipole polarizabilities and compare with those available in the literature.  The error bars in the 
experimental results are large for all the systems and our results lie within 
them. The results of the calculations that are given in the above table are 
obtained using a variety of many-body theories. Our results are in reasonable 
agreement with them except for a few cases where they differ by more than 5\%. 

Mitroy and Bromley \cite{mitroy} have used oscillator strengths from a  
semiempirical approach to obtain these quantities. Some results based on an {\it ab
initio} method that combines the configuration interaction (CI) method
and many-body perturbation theory (MBPT) are available \cite{porsev,porsev1,kozlov}. In
these calculations, the valence electrons correlation effects are evaluated by the CI
method whereas the core electrons correlation effects are calculated using the MBPT method.
However, these calculations consider the core-polarization effects in the 
framework of finite order MBPT, while our CC method takes them into account to all orders.
\begin{table}
\caption{Static dipole polarizabilities in divalent atoms: Be, Mg, Ca, Sr, Ba and Yb (in au).}
\begin{ruledtabular}
\begin{center}
\begin{tabular}{llll}
Atoms &  Expts & Others & This work \\
\hline\\
Be &  & 37.755$^a$, 37.69$^b$, 37.9$^c$  & 37.80 \\
&  & &  \\
Mg & 71.5(3.1)$^d$ & 71.35$^b$, 72.0$^c$, 71.3(7)$^e$ & 73.41 \\
 &  & &  \\
Ca & 169(17)$^d$ &  159.4$^b$, 152.7$^c$, 157.1(1.3)$^e$ & 154.58 \\
 &  & 152$^f$, 158.0$^g$, &  \\
 &  & &  \\
Sr &  186(15)$^d$ & 201.2$^b$, 193.2$^c$, 197.2(2)$^e$ & 199.71 \\
 &  & 190$^f$, 198.9$^g$, &  \\
&  & &  \\
Ba &  268(22)$^h$ & 264$^i$, 273$^f$, 273.9$^g$ & 268.19 \\
 &  & &  \\
Yb &  142(36)$^j$ & 111.3(5)$^e$, 141.7$^k$, 157.30$^l$ & 144.59 \\
\end{tabular}
\end{center}
\end{ruledtabular}
$^a$Reference: \cite{komasa}\\
$^b$Reference: \cite{mitroy}\\
$^c$Reference: \cite{patil}\\
$^d$Reference: \cite{miller2}\\
$^e$Reference: \cite{porsev}\\
$^f$Reference: \cite{andrzej}\\
$^g$Reference: \cite{lim}\\
$^h$Reference: \cite{schwartz}\\
$^i$Reference: \cite{kozlov}\\
$^j$Reference: \cite{miller4}\\
$^k$Reference: \cite{buchachenko}\\
$^l$Reference: \cite{chu}\\
\label{tab1}
\end{table}
Lim and Schwerdtfeger had employed the scalar-relativistic Douglas-Kroll Hamiltonian
 to determine some of these quantities \cite{lim}. In these calculations, they 
had shown the
 importance of the relativistic effects. There are also a few calculations
available for the Yb polarizabilities using the CC method \cite{buchachenko,wang}, where the atomic orbitals are evaluated using the molecular symmetries. Recently TDDFT 
was used for calculating the same quantities  \cite{chu}, but this method treats  
exchange and correlation effects via local potentials. The
main difference between the methods on which all these calculations are based  and ours  is that we 
calculate both the unperturbed and the first order perturbed wavefunctions using a  
RCC approach  that implicitly takes into consideration all the intermediate states.

\begin{table}
\caption{Contributions from DF and important perturbed CC terms (in au) for the dipole polarizabilities.}
\begin{ruledtabular}
\begin{center}
\begin{tabular}{lcccc}
Atoms &  DF & $(\overline{D}T_{1}^{(1)} + cc)-$DF & $\overline{D}T_{2}^{(1)} + cc$ &Norm \\
\hline\\
Be & 45.82 & $-$7.94 & $-$0.09 & 0.02 \\
Mg & 82.44 & $-$8.77 & $-$0.21 & 0.03 \\
Ca & 184.14 & $-$29.23 & $-$0.07 & $-$0.26 \\
Sr & 234.41 & $-$34.46 & $-0.17$ & $-$0.08 \\
Ba & 328.32 & $-$61.18 & 0.09 & 0.81 \\
Yb & 183.32 & $-$39.86 & 0.032 & 1.10  \\
\end{tabular}
\end{center}
\end{ruledtabular}
\label{tab2}
\end{table}

\begin{table}
\caption{Contributions from important $nsmp \ ^3P_1^0$ and $nsmp \ ^1P_1^0$ states in the form of single particle orbitals through the ($\overline{D}T_{1}^{(1)} + cc$) CC terms to the dipole polarizabilities.}
\begin{ruledtabular}
\begin{center}
\begin{tabular}{lcccc}
Atoms &  $ns$ & $mp$ & DF & $\overline{D}T_{1}^{(1)}+cc$ \\
\hline\\
  & &  & & \\
Be & 2s1/2 & 2p1/2 & 1.154 & 0.893 \\
 & 2s1/2 & 3p1/2 & 8.261 & 6.596 \\
 & 2s1/2 & 4p1/2 & 5.529 & 4.583 \\
 & 2s1/2 & 2p3/2 & 2.308 & 1.816 \\
 & 2s1/2 & 3p3/2 & 16.521 & 13.484 \\
 & 2s1/2 & 4p3/2 & 11.059 & 9.390 \\
  & &  & & \\
Mg & 3s1/2 & 3p1/2 & 3.521 & 2.292 \\
 & 3s1/2 & 4p1/2 & 15.701 & 10.279 \\
 & 3s1/2 & 5p1/2 & 7.780 & 5.254 \\
 & 3s1/2 & 3p3/2 & 7.009 & 6.742 \\
 & 3s1/2 & 4p3/2 & 31.351 & 31.308 \\
 & 3s1/2 & 5p3/2 & 15.605 & 17.030 \\
  & &  & & \\
Ca & 4s1/2 & 4p1/2 & 20.550 & 17.790 \\
 & 4s1/2 & 5p1/2 & 32.290 & 29.712 \\
 & 4s1/2 & 6p1/2 & 6.928 & 8.073 \\
 & 4s1/2 & 4p3/2 & 40.358 & 32.079 \\
 & 4s1/2 & 5p3/2 & 64.805 & 53.218 \\
 & 4s1/2 & 6p3/2 & 14.097 & 13.437 \\
  & &  & & \\
Sr & 5s1/2 & 5p1/2 & 30.288 & 25.795 \\
 & 5s1/2 & 6p1/2 & 38.465 & 35.555 \\
 & 5s1/2 & 7p1/2 & 6.838 & 8.402 \\
 & 5s1/2 & 5p3/2 & 56.608 & 45.432 \\
 & 5s1/2 & 6p3/2 & 78.344 & 67.307 \\
 & 5s1/2 & 7p3/2 & 14.665 & 15.247 \\
  & &  & & \\
Ba & 6s1/2 & 6p1/2 & 55.019 & 47.128 \\
 & 6s1/2 & 7p1/2 & 45.796 & 45.569 \\
 & 6s1/2 & 8p1/2 & 4.631 & 8.292 \\
 & 6s1/2 & 6p3/2 & 95.404 & 71.484 \\
 & 6s1/2 & 7p3/2 & 98.613 & 81.924 \\
 & 6s1/2 & 8p3/2 & 11.420 & 11.623 \\
  & &  & & \\
Yb & 6s1/2 & 6p1/2 & 26.578 & 19.824 \\
 & 6s1/2 & 7p1/2 & 29.112 & 24.629 \\
 & 6s1/2 & 8p1/2 & 3.234 & 3.925 \\
 & 6s1/2 & 6p3/2 & 43.498 & 32.953 \\
 & 6s1/2 & 7p3/2 & 61.077 & 52.816 \\
 & 6s1/2 & 8p3/2 & 8.518 & 9.522 \\
\end{tabular}
\end{center}
\end{ruledtabular}
\label{tab5}
\end{table}
\begin{figure}[h]
\includegraphics[width=8.8cm]{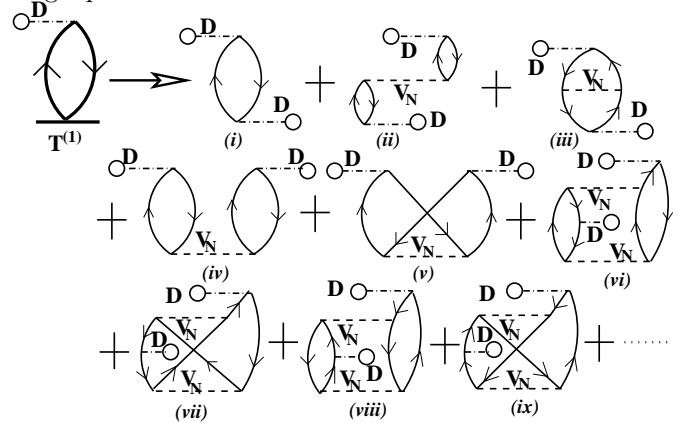}
\caption{Breakdown of the $\overline{D}T_{1}^{(1)}$ CC diagram in terms of 
lower order MBPT and RPA diagrams that contribute significantly to the
polarizability calculations. Here, D and V$_N$ represent the dipole 
and normal order Coulomb interaction operators which are
shown as single dotted and dashed lines, respectively.}
\label{fig1}
\end{figure}
To emphasize the importance of correlation effects in these calculations,
we present the DF and the leading RCC contributions in Table \ref{tab2} 
for the electric dipole polarizabilities. For all
the cases that have been considered, the DF results are larger than the total results.
From the individual RCC contributions, we find that only  
the terms arising from $\overline{D}T_{1}^{(1)}$ and its conjugate ($cc$) are significant. 
Given that these terms include 
the DF, leading core-polarization 
and other important correlation effects to all orders, it is not surprising that they should collectively make up the largest contribution. In Fig \ref{fig1}, we give the breakdown 
of $\overline{D}T_{1}^{(1)}$ in terms of the DF, random phase approximation 
(RPA) and other 
diagrams.  All the above mentioned calculations
where these terms have been evaluated are based on finite order MBPT, 
but in our present method, we treat them to all orders in the residual Coulomb 
interaction. 
The $^3P_1$ and $^1P_1$ configurations are of crucial importance for the leading term of  $\overline{D}T_{1}^{(1)}$ and its conjugate. In Table \ref{tab5}, we present the contributions arising from different combinations of single particle orbitals.  

\section{Conclusion}
We have carried out calculations of electric dipole polarizabilities 
for many alkaline earth atoms and ytterbium using the 
RCC method and  highlighted the importance of the 
correlation effects. The novelty in 
these calculations is that it avoids the sum-over-states 
approach in determining polarizabilities and implicitly considers all the intermediate states by solving the perturbed RCC wavefunction to first order in the dipole and all orders in the residual Coulomb interaction.

\section{Acknowledgment}
We thank Professor A. Dalgarno and Dr. P. Zhang for useful discussions.
The computations reported in this paper were performed using C-DAC's TeraFlop 
Super Computing facility at Bangalore. We thank Mr N. Mohanram, Dr. Subroto 
Chattopadhyay and their colleagues for cooperation.


\begin{thebibliography}{26}
\bibitem{miller1}
T. M. Miller and B. Bederson, Adv. in Atomic and Molecular Phys. {\bf 13}, 1 (1977).
\bibitem{pethick}
C. J. Pethick and H. Smith, {\it Bose-Einstein Condenstaion in Dilute Gases},
Cambridge University Press (2002).
\bibitem{dalgarno}
A. Dalgarno and Lewis, Proc. Roy. Soc. A {\bf 233}, 70 (1955).
\bibitem{dalgarno1}
A. Dalgarno and H. A. J. McIntyre, Proc. Phys. Soc. {\bf 85}, 47 (1965).
\bibitem{kundu}
B. Kundu and D. Mukherjee, Chem. Phys. Lett. {\bf 179}, 468 (1991).
\bibitem{stanton}
J. F. Stanton and R. J. Bartlett, J. Chem. Phys. {\bf 99}, 5178 (1993).
\bibitem{kallay}
M. Kallay and J. Gauss, J. Mol. Struct. (THEOCHEM) {\bf 768}, 71 (2006).
\bibitem{safronova}
U. I. Safronova, W. R. Johnson and M. S. Safronova, Phys. Rev. A {\bf 76}, 042504 (2007).
\bibitem{bijaya1}
B. K. Sahoo, B. P. Das, R. K. Chaudhuri, D. Mukherjee, R. G. E. Timmermans and K. Jungmann, Phys. Rev. A {\bf 76}, 040504(R) (2007)
\bibitem{porsev}
S. G. Porsev and A. Derevianko, Phys. Rev. A {\bf 74}, 020502(R) (2006)
\bibitem{degenhardt}
C. Degenhardt {\it et al.}, Phys. Rev. A {\bf 72}, 062111 (2005)
\bibitem{boyd}
M. M. Boyd, A. D. Ludlow, S. Blatt, S. M. Foreman, T. Ido, T. Zelevinsky and J. Ye, Phys. Rev. Lett. {\bf 98}, 083002 (2007)
\bibitem{barber}
Z. W. Barber, C. W. Hoyt, C. W. Oates, L. Hollberg, A. V. Taichenachev and V. I. Yudin, Phys. Rev. Lett. {\bf 96}, 083002 (2006)
\bibitem{takahashi}
Y. Takahashi et al, {\it Proceedings of CP violation and its origin}, edited by K. Hagiwara, KEK report, pg. 259 (1997).
\bibitem{andrzej}
A. J. Sadlej, M. Urban and O. Gropen, Phys. Rev. A {\bf 44}, 5547 (1991)
\bibitem{chu}
X. Chu, A. Dalgarno and G. C. Groenenboom, Phys. Rev. A {\bf 75}, 032723 (2007)
\bibitem{cizek}
J. Cizek, J. Chem. Phys. {\bf 45}, 4256 (1966)
\bibitem{lindgren}
I. Lindgen and J. Morrison, {\it Atomic Many-Body Theory}, edited by G. Ecker, P. Lambropoulos, and H. Walther ( Springer-Verlag, Berlin, 1985)
\bibitem{mitroy}
J. Mitroy and M. W. J. Bromley, Phys. Rev. A {\bf 68}, 052714 (2003)
\bibitem{porsev1}
S. G. Porsev, Y. G. Rakhlina and M. G. Kozlov, Phys. Rev. A {\bf 60}, 2781 (1999)
\bibitem{kozlov}
M. G. Kozlov and S. G. Porsev, Eur. Phys. J. D {\bf 5}, 59 (1999)
\bibitem{lim}
I. S. Lim and P. Schwerdtfeger, Phys. Rev. A {\bf 70}, 062501 (2004)
\bibitem{buchachenko}
A. A. Buchachenko, M. M. Szczesniak and G. Chalasinski, J. Chem. Phys. {\bf 124}, 114301 (2006)
\bibitem{wang}
Y. Wang and M. Dolg, Theo. Chem. Acc. {\bf 100}, 124 (1998)
\bibitem{komasa}
J. Komasa, Phys. Rev. A {\bf 65}, 012506 (2001)
\bibitem{patil}
S. H. Patil, J. Phys. D {\bf 10}, 341 (2000)
\bibitem{miller2}
T. M. Miller, {\it Atomic and Molecular Polarizabilities}, CRC Press, Boca Raton, Florida {\bf 76}, pp. 10 (1995)
\bibitem{schwartz}
H. L. Schwartz, T. M. Miller and B. Bederson, Phys. Rev. A {\bf 10}, 1924 (1974)
\bibitem{miller4}
T. M. Miller, CRC handbook of chemistry and physics, Edited by D. R. Lide 77th edition CRC, Boca Raton (1996)

\end{thebibliography}
\end{document}